\documentclass[apj]{emulateapj}
\usepackage{graphicx,url,amssymb,amsmath,longtable,rotating,units,wasysym,listings,bm}
\usepackage{natbib,verbatim,enumerate}

\def\apjl{ApJL }

\def\apj{ApJ }

\def\araa{ARAA }
\def\aap{A\&A }
\def\nat{Nature }
\def\mnras{MNRAS }
\def\prd{Phys. Rev. D. }

\def\llgrb{{\it ll}GRB }
\def\llgrbs{{\it ll}GRBs }
\def\llgrbns{{\it ll}GRB}
\def\llgrbsns{{\it ll}GRBs}

\begin{document}
\normalsize

\author{Ehud Nakar$^1$}

\altaffiltext{1}{Raymond and Beverly Sackler School of Physics and Astronomy, Tel Aviv University, Tel Aviv 69978, Israel}

\title{A unified picture for low-luminosity and long gamma-ray bursts based on the extended progenitor of \llgrb 060218/SN 2006aj}

\begin{abstract}
The relation  between long  gamma-ray bursts  (LGRBs) and  low-luminosity GRBs
(\llgrbsns) is  a long standing puzzle  -- on the one  hand their high energy emission  properties   are  fundamentally  different,  implying   a  different
gamma-ray source, yet both are associated  with similar supernovae of the same
peculiar type (broad-line Ic), pointing at  a similar progenitor and a similar
explosion  mechanism.  Here  we  analyze   the  multi-wavelength  data  of  the
particularly well-observed  SN 2006aj, associated with  \llgrb 060218, finding
that  its  progenitor star  is  sheathed  in  an extended  ($>100R_\odot$),
low-mass ($\sim 0.01M_\odot$) envelope. This  progenitor structure
implies that  the gamma-ray emission in  this \llgrb is generated  by a mildly
relativistic shock breakout. It also suggest a unified picture for \llgrbs and
LGRBs,  where the  key difference  is the  existence of  an extended  low-mass
envelope in \llgrbs and its absence  in LGRBs. The same engine, which launches
a relativistic  jet, can  drive the  two explosions, but,  while in  LGRBs the
ultra-relativistic jet emerges from the  bare progenitor star and produces the
observed gamma-rays,  in \llgrbs  the extended envelope  smothers the  jet and
prevents  the generation  of a  large gamma-ray  luminosity. Instead,  the jet
deposits all its  energy in the envelope, driving a  mildly relativistic shock
that upon  breakout produces a \llgrbns. In addition for  giving a unified
view of  the two  phenomena, this  model provides a natural explanation  to many
observed properties of  \llgrbsns. It also implies that  \llgrbs are a viable
source of the observed extra-galactic diffuse  neutrino flux and that they are
promising    sources     for    future    gravitational     wave    detectors.
\end{abstract}


\section{Introduction}
The gamma-ray emission of LGRBs and  \llgrbs show almost no similarities apart
for   being   detected  by   the   same   instruments.  LGRBs   are   luminous
($10^{50}-10^{52}$  erg/s),  hard   ($\gtrsim  100$  keV),  highly
variable and narrowly collimated with a typical duration of 10-100 s \citep[][
and references therein]{Piran04}. \llgrbs are  fainter by about four orders of
magnitude ($10^{46}-10^{48}$  erg/s), relatively soft ($\lesssim  100$ keV), not highly beamed and
show no significant temporal variability  over their entire duration, which is
often  longer than  1000 s  \citep{Kulkarni98,Campana06,Soderberg06,Kaneko07}. 

To date there  are only four well observed \llgrbsns \footnote{ These are GRBs with luminosity $\lesssim 10^{48}~$erg/s for which an associated SNe was observed - llGRB/SN: 980425/1998bw, 031203/2003lw, 060218/2006aj and 100316D/2010bh.},  compared to hundreds of
LGRBs. However,  this is a  result of their  low luminosity, which  limits the
detection to a  distance of $\sim 100$  Mpc, compared to LGRBs  which are seen
through the entire Universe. In fact, \llgrbs are much more common than LGRBs,
and are the most abundant known relativistic explosions in the nearby Universe
\citep{Soderberg06,Pian06}. Thus,  \llgrbs are  of special interest,  both for
the understanding  of GRBs and their connection to SNe, and  as sources of high  energy non-electromagnetic
signals,  such   as  gravitational-waves  \citep[e.g.,][]{Kotake12,Birnholtz13,Ando13},
neutrinos  \citep[e.g.,][]{Murase06,Murase13}   and  cosmic-rays\citep[e.g.,][]{Budnik08,Liu11}.

Based on high  energy emission alone there  is no reason to  assume that LGRBs
and \llgrbs  are related. Moreover,  theoretical considerations show  that the
gamma-rays seen  in \llgrbs cannot be  produced in the same  environment where
the gamma-rays  in LGRBs  are generated  \citep{Bromberg11a}. It  is therefore
puzzling that  these two  apparently different GRB  types are  both associated
with very similar peculiar SNe of the  rare broad-line Ic type \citep[e.g.,][]{Melandri14}. These SNe show
no  signs of  H or  He  in their  spectra,  an indication  of highly  stripped
progenitors.   Their   ejecta  have   unusually   high   velocities  for   SNe
(10,000$-$30,000 km/s),  their peak  luminosities indicate a  relatively large
amount of synthesized $^{56}$Ni, and the  total kinetic energy carried by some
of these  SNe is  unusually high \citep[][and  references therein]{Woosley06}.
The similarity  of the  associated SNe  suggests that  \llgrbs and  LGRBs have
similar  progenitors  and  similar  inner  explosion  mechanism.  The  natural
question  that  arises  is  how  similar  explosions  produce  such  different
gamma-ray                                                             signals.

Here we  approach this puzzle  by analyzing  the early (first  day) optical/UV
light curve of SN 2006aj, which is  associated with \llgrb 060218, in order to
study its  progenitor structure.  \llgrb 060218/SN 2006aj  has the  best early
observational  coverage  out of  the  four  well  observed \llgrbs  and  their
associated SNe. It includes Swift  continuous gamma-ray, X-ray, UV and optical
observations ranging from $10^2-10^6$ s after the explosion \citep{Campana06},
many  optical  spectra  starting  less  than two  days  after  the  explosion
\citep{Mirabal06,Modjaz06,Mazzali06,Pian06,Sollerman06} and radio observations
starting a day after the explosion  \citep{Soderberg06}. In fact SN 2006aj has
probably the  most detailed early  optical/UV photometric coverage out  of the
thousands SNe  observed to date.  

The unique feature of the optical/UV light curve of SN 2006aj is that it shows
two peaks. Using a recently developed method for the analysis of double-peaked
SNe  \citep{NakarPiro14}   we  constrain  the  progenitor   properties.  These
properties are then  used to learn about  the physics of \llgrbs  and on their
relation                               to                               LGRBs.

The paper  structure is as  follows. Section \S\ref{sec:Progenitor} presents the analysis of the
optical/UV  light  curve and the resulting constraints on  the  progenitor structure  of  \llgrb
060218/SN 2006aj.  These constraints strongly
support  the suggestion  that \llgrbs  are generated  by shock  breakouts (\S\ref{sec:Breakout}). 
A unified  picture for \llgrbs and  LGRBs that
naturally  explain   the  similarities   and  differences  between   them is presented in \S\ref{sec:Unified}.  This picture provides  a simple explanation
to  the unique  velocity  profile  of SNe  associated  with  {\it ll}GRB (\S\ref{sec:Velocity}).  The
implication  of  this picture  for  \llgrbsns'  neutrino and  gravitational  wave
emission   is   discussed   in    \S\ref{sec:nonthermal}.

\section{The progenitor of \llgrb 060218/SN 2006aj}\label{sec:Progenitor}

\begin{figure}[!t]
\includegraphics[width=1\columnwidth]{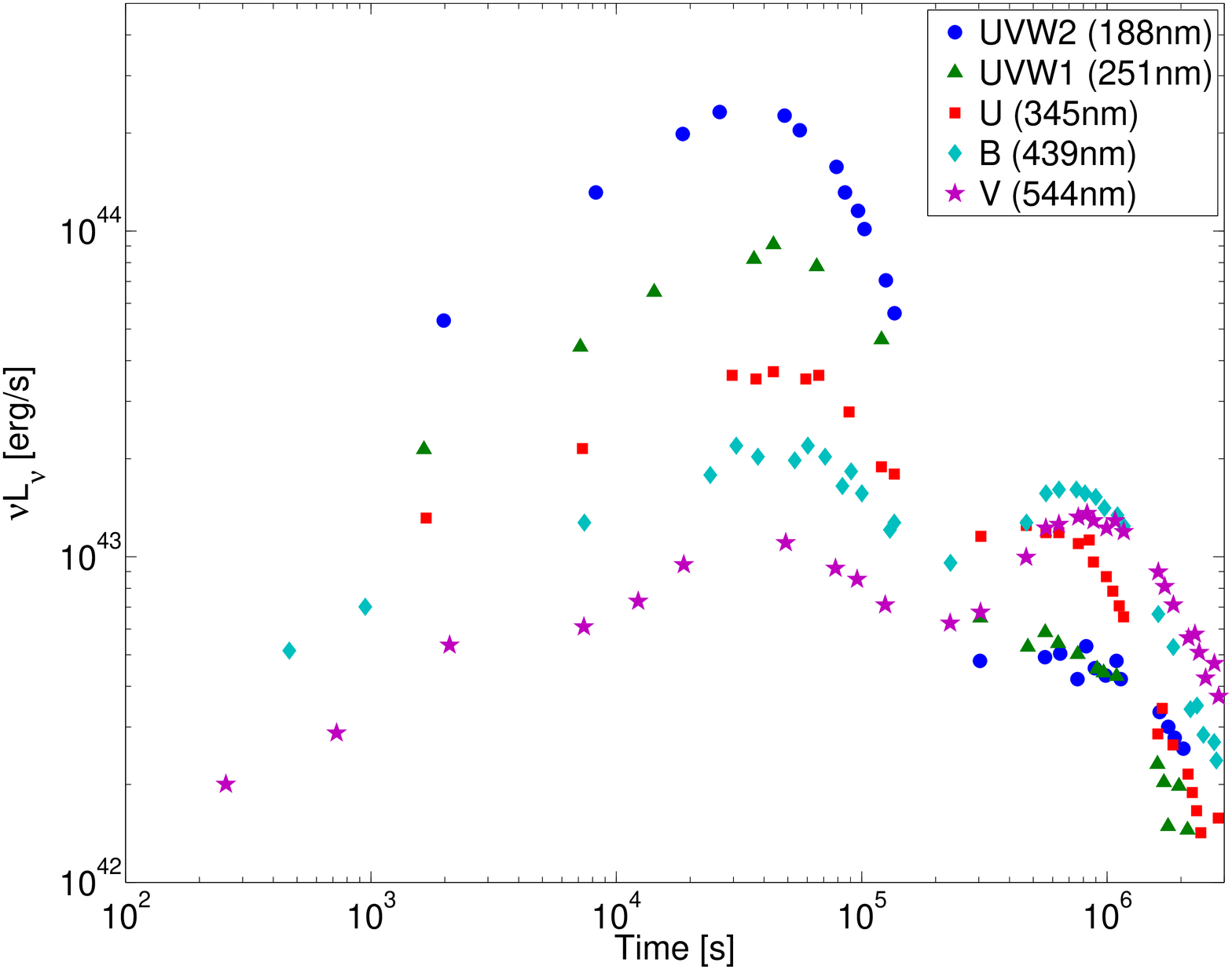}
\caption{Swift UVOT light curve of SN 2006aj in five filters (central wavelength given in the legend) from \cite{Campana06}. Following \cite{Campana06} the light curve is corrected for Galactic reddening, $E(B - V) = 0.14$  (assuming a Galactic reddening curve with $R_V=3.1$), and a host galaxy reddening $E(B - V) = 0.20$ (assuming a Small Magellanic Cloud reddening curve).}
\label{fig:UVOT_lc}%
\end{figure}

Figure  \ref{fig:UVOT_lc} depicts  the  optical/UV light  curve  of SN  2006aj
as taken from \cite{Campana06}. It shows two peaks in  the optical bands, at $t \approx 10$
hr  and  $t\approx  10$  days, where $t$  is time  since  first  detection  of  the
gamma-rays, estimated  here as the  explosion time. Such  double-peaked light
curves          are           very          rare           among          SNe.
In typical  SNe  the light  curve is dominated  by one  of two
power sources:  (i) the internal  energy deposited by  the SN shock,  known as
``cooling envelope emission", or (ii) the radioactive decay of $^{56}$Ni. Each
one of  these power sources  produces only a single  peak in the  optical and, in typical SNe, the time scales of the maximal contribution to the optical light from each of the two sources are comparable. Therefore, observed  SN light  curves usually contain  only a  single optical peak which is powered by the stronger power source at any given SNe. This is cooling envelope in explosions of red supergiants, such as type II-P SNe, and $^{56}$Ni in explosions of more compact progenitors, such as type I and 1987-like SNe.

Two peaks are observed in rare cases where at early time the emission is powered by the
cooling envelope, which then decays sharply on a time scale comparable to that
of  the rising  $^{56}$Ni  contribution. This  behavior  requires an  atypical
progenitor structure  of a compact massive  core that is engulfed  by extended
low-mass material  \citep{Hoflich93,Bersten12,NakarPiro14}. The second  peak in
these cases is  similar to the main  peak of a typical  $^{56}$Ni powered SNe,
and thus its  properties provide an estimate  of the total ejecta  mass and of
the  $^{56}$Ni mass.  In the  case of  SN 2006aj  the second  peak is  clearly
powered by $^{56}$Ni and it shows a total ejecta mass of $\sim 2~M_\odot$, out
of which $\sim  0.2~M_\odot$ are $^{56}$Ni \citep{Mazzali06}.  The only natural
source of  the first peak  in SN  2006aj is the  cooling envelope phase  of an
extended  mass  (see appendix \ref{app:firstPeak})  and  thus its  properties  can
provide a robust estimate of the radius and the mass of the extended material.
Here the  results of \cite{NakarPiro14}  are used to derive  these constraints.

The mass of the extended material can be estimated from the time of the first peak, $t_p$ \citep{NakarPiro14}:
\begin{equation}
	M_{ext} \approx  0.01  \frac{v_{ext}}{~0.2\,c}
	\left( \frac{t_{p}}{10\,{\rm hr}}\right)^2~M_\odot
\end{equation}
where  $c$ is  the light  speed and  $v_{ext}$ is  the velocity  to which  the
extended material is accelerated  by the explosion. Spectroscopic observations
limit  $v_{ext}>0.1$  c,  the   measured  photospheric  velocity  at  $t=2.89$
day \citep{Pian06}. On  the high  end it  is most  likely that  $v_{ext}<0.3$ c,
since at  higher velocity the  kinetic energy  carried by $M_{ext}$  would be larger
than the kinetic energy deposited by  the explosion in the massive core ($\sim
10^{51}$  erg;  \citealt{Mazzali06}). 

The  pre-explosion  radius  of the  extended
material, $R_{ext}$, can  be estimated by the bolometric luminosity
at the first peak.  The colors before and during the first  peak are very blue
and they  are constant  in time  \citep{Simon10}, as  expected if  the observed
bands are on the Rayleigh-Jeans part  of a blackbody spectrum during that time
(i.e., with temperature $T(t  \leq t_p) \gtrsim  50,000$ K;  see consistency check  below). This
implies that the  total luminosity seen in  UV during the peak,  $\sim 3 \cdot
10^{44}$ erg, is  only a lower limit on the  true bolometric luminosity, which
may  be significantly  higher. Since  $R_{ext}$  is linear  in the  bolometric
luminosity \citep{NakarPiro14}, the  available observations  set a  lower limit:
\begin{equation}
	R_{ext}  \gtrsim 10^{13} 
	 \left(\frac{v_{ext}}{~0.2~c}\right)^{-2}{\rm cm}
\end{equation}      
This is consistent with  the lack of color evolution at  $t<t_p$ and the model
prediction  that temperature  is  dropping  with time,  reaching  at the  peak
$T(t_p)     \approx    50,000     (R_{ext}/10^{13}\,{\rm    cm})^{1/4}{\rm~K}$
\citep{NakarPiro14}. Thus,  the optical/UV light  curve of SN  2006aj indicates
that its progenitor had a relatively compact  core of several solar masses, surrounded by
$\sim  0.01~M_\odot$ which  is  extended to  a radius  of  a supergiant.  This
structure is very different than the typically expected structure of a fully H
stripped progenitor,  based on stellar evolution  models, yet it must  be very
common in GRB  progenitors given that \llgrbs are more  common than LGRBs. This  progenitor structure has  several far
reaching  implications for  the physics  of \llgrbs  and their  associated SNe, which are discussed in the following sections.

\section{Shock breakout origin for \llgrbs}\label{sec:Breakout}

The  Thomson optical  depth  of the  extended material  is  high, $\sim  3,000
(R_{ext}/10^{13}\,{\rm cm})^{-2}$. As a result, the breakout of  the shock
driven by the explosion takes place at $R_{ext}$. Radio observations show that
the leading  edge of  the outflow  is mildly  relativistic \citep{Soderberg06}, implying  that the
breakout must  be at least  at a  mildly relativistic velocity,  i.e., $v_{bo}
\gtrsim 0.5$ c. Since rate considerations indicate that the gamma-rays in \llgrbs are not
strongly  beamed \citep{Soderberg06} and  late SN  spectroscopy and  polarimetry
show no  signs of  ejecta a-sphericity \citep{Mazzali07},  the breakout  is not
expect to  strongly deviate  from a spherical  symmetry. In  that case  the main
characteristics  of   a  mildly   relativistic  shock  breakout   signal,  its
luminosity, duration  and typical photon  energy, depend only on  the breakout
radius                                                            \citep{NS12}:
\begin{equation*}
	L_{bo} \sim 2 \cdot 10^{46} \frac{R_{ext}}{3 \cdot 10^{13}\,{\rm cm}} ~{\rm erg~s^{-1}}	
\end{equation*}
\begin{equation}\label{eq:breakout}
	t_{bo} \sim 1000 \frac{R_{ext}}{3 \cdot 10^{13}\,{\rm cm}} ~{\rm s~}
\end{equation}
\begin{equation*}
	T_{bo} \sim 50 ~{\rm keV}
\end{equation*}
This  is  similar  to  the  actual gamma-ray  signal  of  \llgrb  060218  where
$L_{bo,obs}  \approx 3  \cdot 10^{46}~{\rm  erg~s^{-1}}$, $t_{bo,obs}  \approx
1,000$ s  and $T_{bo,obs} \approx  40$ keV  \citep{Kaneko07} and it  fits very
well  to a  breakout  radius $R_{ext}  \sim  3 \cdot  10^{13}$  cm. Thus,  the
combination of optical/UV  and radio observations imply that  a shock breakout
signal  is inevitable  and that  its properties  are similar  to the  observed
\llgrb. As shock breakout also explains a large range of properties of the high
energy emission  from \llgrbs (e.g., smooth  profile, spectral  evolution, low
beaming;  \citealt{NS12}),  this  result  practically  implies  that  the  entire
gamma-ray signal in  \llgrb 060218 is generated by a  mildly relativistic shock
breakout, without  any significant  contribution from  a relativistic  jet. It
also lends a very strong support for  the suggestion that all \llgrbs are shock
breakouts        \citep{Kulkarni98,Tan01,Campana06,Waxman07,Li07,Katz10,NS12}.

\section{A unified picture for LGRBs and \llgrbs}\label{sec:Unified}
If all \llgrb progenitors have a similar structure to that of \llgrb 060218 then it provides a natural solution to the puzzle why two explosions
with  similar inner  workings produce such different  gamma-ray signals.
According to the standard model for LGRBs the burst
is   powered  by   a   central  engine   that   launches  a highly   collimated
ultra-relativistic bipolar jet. In order to  produce a LGRB the jet must first
punch its  way through the star  and then expand freely  at ultra-relativistic
velocities  to radii where generated  gamma-rays can  be seen  by the
observer. While the jet drills through the dense stellar matter
its energy is  dissipated and the engine must continue to supply power into
the jet if it is to succeed punching through the star and produce the observed
LGRB   \citep{Zhang03,Morsony07,Mizuta09,Bromberg11b}.    Thus,   a   necessary
condition for the production of a LGRB is that the engine working time is long enough to allow the  jet  to drill  through  the star.  Observations
indicate  that a typical  LGRB engine  launches  a jet at  a typical  isotropic
equivalent luminosity  of $L_{iso} \sim  10^{51}$ erg/s and a  typical opening
angle $\theta_j \sim  10^o$  over a  typical duration of $\sim  20$ s \citep{Piran04}. The
total energy carried by the jet, after correction  for beaming, is $\sim
10^{51}$ erg.  If the progenitor is a bare  H stripped  star of  several solar masses and several solar  radii it takes $\sim 10$ s for  the jet to penetrate
through  the star  (see appendix \ref{app:jets}; \citealt{Bromberg11b}), implying  that the  jet can
successfully emerge from  the star and that the collapse  of such a progenitor
can lead  to a LGRB.  

\begin{figure}[!t]
\includegraphics[width=1\columnwidth]{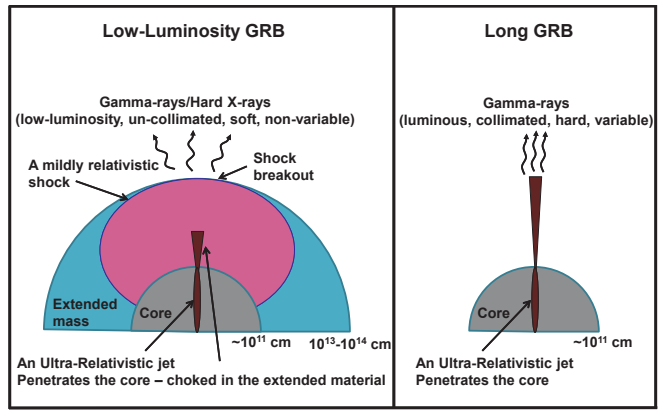}
\caption{A  schematic  sketch  illustrating  the  similarity  and  differences
between \llgrbs and LGRBs. Both explosions  go through a collapse of a similar
core which leads to the formation of a  similar GRB engine and to a similar SN
explosion. In both  types the GRB engine  launches ultra-relativistic narrowly
collimated jet, which penetrates through the core. In LGRBs the jet is free to
expand as soon  as it is out of  the core where it produces  a luminous, hard,
narrowly collimated  beam of gamma-rays which  can vary in time  on short time
scales. In  \llgrb the jet  emerges from the  core into the  low-mass extended
material where it is choked and any radiation that it produces is absorbed and
cannot reach  to the  observer. The  jet energy is  deposited in  the extended
material driving a  strong shock into it. The shock  is much less relativistic
than the jet  (most likely Newtonian) and it accelerates  before breakout (often
to  a   mildly  relativistic  velocity).  Upon breakout it   produces  low-luminosity  soft
gamma-rays  which  show no  significant  variability  with  time and  are  not
narrowly                                                              beamed.}
\label{fig:sketch}%
\end{figure}

The picture,  however, is very  different if  there is an  additional extended
envelope surrounding the
massive  core, similar  to the one  found here  for \llgrb 060218. Although the  extended material  mass is low  its large
radius makes it  very hard for a jet  to penetrate. In fact the  time that the
engine must  work in order  for the jet to  drill through the  entire extended
mass  is  (appendix  \ref{app:jets}):  \begin{equation}  t_{eng}  \gtrsim  150
\left(\frac{L_{iso}}{10^{51}        {\rm erg/s}}\right)^{-\frac{1}{2}}
\left(\frac{R_{ext}}{3\times10^{13}       {\rm       cm}}\right)^{\frac{1}{2}}
\left(\frac{M_{ext}}{10^{-2}       M_\odot}\right)^{\frac{1}{2}}      {\rm s}.
\end{equation} This time is considerably  longer than the typical working time
of a LGRB engine. Thus, a collapse  of the progenitor of \llgrb 060218 and the
formation of  a LGRB engine at its center will not lead to  an observed LGRB.  Instead, the
launched  jet,  which penetrates  the  stellar  core,  is choked  while  still
propagating in the extended material.  The energy  carried by  the  jet ($E_{jet} \sim 10^{51}$ erg) is  then  deposited  in the  extended  mass accelerating it to high velocity ($v_{ext}\approx 0.3 c \left[\frac{E_{jet}}{10^{51}{\rm\,erg}}\right]^{\frac{1}{2}}\left[\frac{M_{ext}}{10^{-2} M_\odot}\right]^{-\frac{1}{2}} $) and driving into it a strong shock. The shock accelerates further 
at the dropping  density gradient near $R_{ext}$ and upon  breakout produces a
\llgrbns. 

Note that while the energy deposition is done by  a narrow jet and
is  therefore highly  aspherical, the shock upon breakout can be quasi-spherical. The reason is that the jet  is  choked  long before  it
approaches $R_{ext}$ and  the resulting  blast wave  becomes much  more spherical during its propagation  before it breaks out at $R_{ext}$.  A schematic sketch
of the similarities  and differences between \llgrbs and  LGRBs according this
picture       is      illustrated       in      figure       \ref{fig:sketch}.

\section{The uncommon velocity profile of SNe associated with \llgrbs}\label{sec:Velocity}
This  scenario resolves  yet another  puzzle  related to  SNe associated  with
\llgrbs -- why is the kinetic energy in their fast moving ejecta is so high compared to other SNe \citep{Soderberg06}.  In typical  SNe  the
explosion energy is all deposited at the center of the  progenitor. This drives a
shock that crosses  first the bulk of  the mass and then  accelerates at the sharp density drop near the
stellar edge. This acceleration  dictates a
certain relation between the kinetic energy carried by slow and by fast moving
material, where $E_k(v) \propto v^{-5}$ \citep{Sakurai60,MatznerMcKee99}. This relation is seen in regular SNe, but not in \llgrbs where the fast moving ejecta carries much more energy than it predicts \citep{Soderberg06}. For
example, in SN 2006aj about $2 \times  10^{51}$ erg are carried by the bulk of
the mass  at $\sim 10,000$ km/s \citep[]{Mazzali06}.  In a regular SNe  if a mildly relativistic ($>150,000$
km/s)  ejecta exist,  the relation  $E_k(v)  \propto v^{-5}$  dictates that  
it should  carry  $\sim  2 \times 10^{45}$  erg.  Instead,  in   SN  2006aj  radio
observations indicate  that the mildly relativistic  material carries $\gtrsim
10^{48}$ erg  \citep{Soderberg06,BarniolDuran14}. This observed property  of SN
2006aj is naturally explained by the picture  of \llgrbs presented here. In
this picture  the energy in  the slow moving material  is deposited by  the SN
explosion mechanism  at the center, while  the observed $>10^{48}$ erg  in the
fast moving  material is deposited directly  by a GRB jet,  thereby decoupling
the  amount   of   energy   carried  by   each   of   the   components.

\section{Neutrinos  and  gravitational  waves   from  \llgrbs}  \label{sec:nonthermal}
An intriguing implication  of the arising picture is the  prospects for future
detection of  non-electromagnetic signal from \llgrbsns.  The extreme energies
and velocities involved in LGRB engines  and jets make them a potential source
of gravitational  waves (GW), neutrinos  and high-energy cosmic  rays. However
LGRBs are very rare in the local Universe and are typically seen at a distance
$>$Gpc. Here  we suggest that \llgrbs  harbor the same engine  as LGRBs, which
produces similar ultra-relativistic jets. The  propagation of an \llgrb jet is
similar to that  of a LGRB jet  within the massive core. After  the jet breaks
out of the core and into the  \llgrb extended envelope the envelope density is
low, so  the pressure  in the  cocoon does  not affect  the jet  (see appendix
\ref{app:jets}). Thus, at any  location that is far from the  jet head the jet
is unaware  of the  extended envelope. Therefore,  all the  physical processes
that take place during  the formation of the engine, the  launching of the jet
and the jet  propagation in LGRB also  take place in \llgrbs up  to the radius
where \llgrb jets are choked  in their progenitors' extended envelopes, namely
$\sim 10^{12}-10^{13}$ cm. Thus, the same emission generated by a LGRB engine
and by its jet while it propagates up to a radius of $\sim 10^{12}-10^{13}$ cm
are expected  to be generated also  by a \llgrb. This  includes photons, high energy particles (cosmic-rays), neutrinos and  GWs. The extended envelope has
a  Thompson optical  depth $\gg  100$  and therefore  photons and  cosmic-rays
cannot  escape  through the  extended  envelope  (the cross-section  for  $pp$
inelastic collision  at $\sim$ TeV  energies is $\sim 0.1$  Thomson). However,
the envelope is transparent to neutrinos and  GWs. It is  therefore worth
considering the  implications of the model  presented here for neutrino  and GW
emission  from \llgrbsns,  especially given  that \llgrbs  are more  common than
LGRBs.

LGRBs  are expected  to  be bright  sources of  high  energy neutrinos  ($\sim
10^{14}-10^{16}$  eV). The  most  promising production  site  of neutrinos  is
internal shocks within  the relativistic jet \citep{Waxman97}.  At radii that
are large enough  $>10^{11}-10^{12}$ cm these shocks are  collisionless and are
therefore expected  to efficiently  accelerate protons  (at smaller  radii the
shocks  are  radiation mediated  and  no  efficient particle  acceleration  is
expected;  \citealt{Levinson08}). At  radii that are small  enough, $\lesssim
10^{14}$, the photon  density is high enough to allow  an efficient photo-pion
production  and   thus  a  generation   of  high  energy   neutrinos.  Current
measurements limit  the neutrino  flux from LGRBs  to $E_0^2\phi_0  \lesssim 2
\times 10^{−10} {\rm~GeV~cm^{-2}~s^{-1}~sr^{-1}}$  per flavor, where $E_0=100$
TeV \citep{IceCube14b}. This upper limit  is comparable to recent estimates of
the  flux expected  from LGRBs  if  particles are  accelerated efficiently  in
internal shocks at radii of  $\sim 10^{12}-10^{14}$ cm \citep{Hummer12} and it
is about two order of magnitude  lower than the measured diffuse neutrino flux
\citep{IceCube14a}.

In  the picture  presented here  \llgrbs are  expected to  be a  much stronger
source of diffuse  neutrino flux than LGRBs\footnote{Note that in the model discussed here the neutrinos are generated in a different environment than in the model discussed by \cite{Murase13}, where they assume that \llgrbs are generated by internal shocks in a wide mildly-relativistic low-luminosity jet. Our findings on the progenitor structure of \llgrb 060218 strongly disfavor the model used by \cite{Murase13}}. First \llgrbs  are more numerous.
The local rate  of \llgrbs, without correction for beaming,  is $\sim 3 \times
10^{-7} {\rm~Mpc^{-3}~yr^{-1}}$,  where beaming  correction is expected  to be
relatively small, $<10$ \citep{Soderberg06,Pian06}. This is compared to a LGRB
local rate, uncorrected for  beaming, of $\sim 10^{-9} {\rm~Mpc^{-3}~yr^{-1}}$
\citep{Wanderman10}. Beaming  correction increases the  true rate of  LGRBs by
about two orders  of magnitude. Thus, \llgrbs are more  frequent than LGRBs by
about an order of magnitude. Second,  The neutrino production of LGRBs depends
on the fraction of high-energy protons that produce pions via interaction with
the observed gamma-rays.  This fraction, denoted as  $f_\pi$, depends strongly
on the burst  parameters and it vary between bursts.  Under optimal conditions
the estimates  are $f_\pi \lesssim 0.1$  \citep{Waxman97,Hummer12}. In \llgrbs
however,  the  jet  is buried  in  an  envelope  that  is optically  thick  to
high-energy protons.  Thus, energy of  protons that  in LGRBs would  have been
released  to the  host galaxy  as cosmic-rays  is converted  in large  part to
neutrinos via  $pp$ interactions. Thus, assuming  that a large fraction  of the
LGRBs' neutrinos are  generated at radii smaller than $\sim  10^{13}$ cm, \llgrbs
are more  efficient in  producing $\sim  100$ TeV neutrinos  by about two orders of magnitude, and possibly more (this is the product of the \llgrb to  LGRB rate ratio and $1/f_\pi$).

As high-energy  protons are accelerate within  the ultra-relativistic narrowly
collimated jets, the  neutrino signal is narrowly beamed as well.  Since the gamma-ray
emission from  \llgrbs is not highly  beamed, most of observed  bursts are not
expected to be accompanied by a neutrino signal. Thus, \llgrbs will contribute
to the  diffuse flux  but they are  not suitable for  a targeted  point-sources
search, similar to the search  conducted for LGRBs \citep{IceCube14a}. Can \llgrbs be then the main source  of the
observed diffuse  flux? \cite{Ahlers14} find  that the sources of  the diffuse
neutrino  flux  produce  a   total  energy  output  of   $\sim  10^{43}
{\rm~erg~Mpc^{-3}~yr^{-1}}$ in  $\sim 100$ TeV neutrinos  and their volumetric
rate,      assuming     transient      sources,      must     be      $\gtrsim
10^{-8}{\rm~Mpc^{-3}~yr^{-1}}$  (as   inferred  from  the  lack   of  neutrino
clustering). Assuming  that each  \llgrb harbor  a relativistic  jet with a typical
energy of  $\sim 10^{51}$ erg  the total energy output  in such jets is  $\sim 3
\times  10^{44}  {\rm~erg~Mpc^{-3}~yr^{-1}}$. Thus,  if  $\sim  10\%$ of  this
energy is converted to high-energy protons  before the jet is choked (i.e., at
radii $\lesssim 10^{13}$  cm) then \llgrbs are producing  the observed diffuse
flux. Assuming that the  typical jet angle is $\sim 10^o$  the rate of \llgrbs
for  which the  neutrino beam  is pointed  towards Earth  is $\sim  0.5 \times
10^{-8}{\rm~Mpc^{-3}~yr^{-1}}$   consistent   with   the   limit   derived   by
\cite{Ahlers14}. Thus, \llgrbs are certainly viable candidates for the origin of the observed extra-galactic neutrino flux!

Finally, if \llgrbs harbor the same engine and relativistic jets as LGRBs then
they  produce  the  same  GW  signals  \citep[e.g.,][]{Kotake12,Birnholtz13}.  A
difference is that while LGRBs are always  observed close to the jet axis, the
line-of-sight to \llgrbs is typically  away from that axis. The GW signal
from the engine can be slightly stronger along the jet axis (up to a factor of
1.6  compared  to  an  average  line-of-sight  observer),  if  its  origin  is
quadrupole  mass inhomogeneity  in a  rotating disk  \citep{Kochanek93}. Other
axisymmetric  GW sources  in the  engine, such  as mass  motions and  neutrino
emission,   vanish  along   the   axis   and  are   strong   at  the   equator
\citep{Kotake12}. The signal from the jet acceleration is also anti-beamed and
is strongest along the equator \citep{Birnholtz13}. Thus, the off-axis viewing
angle of typical  \llgrbs is probably an advantage for  GW detection. The main
advantage of \llgrbs is their much higher rate. The volumetric rate of \llgrbs
is larger  by about an  order of magnitude than  that of all  LGRBs, including
those that  are unobservable since their  gamma-ray beam points away  from the
Earth. If only  LGRBs that points towards earth are  considered then the \llgrb
rate is higher  by almost three order of magnitude. This  is important since targeted
GW search for GRBs (e.g., following  a detection of gamma-rays) is
more sensitive  than a blind search,  increasing the detection  volume by a
factor  of $\approx  3$ \citep{Kochanek93}.  The various  predicted GW  signals from  the
engine and the  jet are expected to be  detectable by future gravitational
wave detectors up to a distance of $\sim 100$ Mpc. The rate of  LGRBs at that  distance is too
low to allow a reasonable probability  for detection. However,
about one \llgrb take place every year within a distance of 100 Mpc, making it a
promising        GW         source        for         future        detectors.

\section{Conclusions}\label{sec:Conclusions}

This paper analyzes the first day optical/UV light curve of SN 2006aj/\llgrb 060218 finding that its progenitor has a compact core engulfed by an extended low-mass material. When the information on this structure is combined with the high velocities inferred from the radio emitting material, it implies that the shock breakout form the extended material must produce a gamma-ray signal that is consistent with the observed \llgrb gamma-ray emission. This indicates that the gamma-rays in 2006aj/\llgrb 060218 are generated by a mildly relativistic shock breakout and it strongly supports the suggestion that the origin of all \llgrbsns ' high energy emission is a shock breakout. 

These results, which are directly based on the observations of SN 2006aj, naturally suggest a picture that unifies LGRBs and \llgrbsns, explaining how two types of explosions that are so different in their gamma-ray signature produce very similar SNe. In this picture  LGRBs and \llgrbs are two manifestations of a similar core collapse process that leads to a similar SN explosion mechanism and a similar GRB central engine, where the observational outcome depends only on the slight differences in the existence, or the lack of, a low-mass extended envelope. This model also provides a simple explanation to the peculiar velocity profile seen in SNe that are associated with \llgrbsns. It also implies that \llgrbs are more promising sources of high energy neutrinos and GWs than LGRBs. \llgrbs are viable candidates as the source of the observed extra-galactic diffuse neutrino flux and are promising GW sources for the next generation GW detectors.

A final comment on  the progenitor structure of SN 2006aj.  While the SN light
curve constrains  $M_{ext}$ and  $R_{ext}$ it does  not strongly  constrain its
density profile. The only  requirement is that most of the
mass $M_{ext}$  is concentrated  near $R_{ext}$ \citep{NakarPiro14}. This  is consistent  with any
density  profile  $\rho(r)$   where  $\rho  r^3$  increases   with  radius  at
$r<R_{ext}$ and  decreases at $r>R_{ext}$.  Thus, we cannot  determine whether
the inferred progenitor structure is in hydrostatic equilibrium or whether the
extended  material  was thrown  out  to  $R_{ext}$  a  short time  before  the
explosion. The  latter option  may be  more attractive  given that  no current
stellar evolution  model predicts a  hydrostatic structure similar to  that SN
2006aj for a fully H stripped star,  while recently there are several lines of
evidence that massive stars go through a strongly enhanced mass-loss episode a
short  time  before   they  explode  \citep{Ofek13,Ofek14,Gal-yam14,Svirski14}.  Yet,
another intriguing speculation is that the progenitor is affected by a binary,
or maybe even by a binary merger (e.g., along similar lines to those suggested
by    \citealt{Chevalier12}),  that put the extended material at place a    short    time    before   the    explosion.

\acknowledgments
This research was partially supported by an ERC starting grant (GRB/SN), ISF grant (1277/13), ISA grant and by the I-CORE Program (1829/12).
I thank Amir Levinson, Dan Maoz, Tsvi Piran, Dovi Poznanski and Amiel Sternberg for enlightening discussions.

%\bibliographystyle{apj}
%\bibliography{2006aj}

%XXXXXXXXXXXXXXXXXXXXXXXXXXXXXXXXXXXXXXXXXXXXXXXXXXXXXXXXXXXXXXXXXXX

		% APPENDICES
												
%XXXXXXXXXXXXXXXXXXXXXXXXXXXXXXXXXXXXXXXXXXXXXXXXXXXXXXXXXXXXXXXXXXX
\appendix 

\section{The power source of the first optical peak} \label{app:firstPeak}
The  analysis  of  the  progenitor  structure  of  SN  2006aj  relies  on  the
identification of  the first optical/UV  peak as a cooling  envelope emission.
Here  this identification  is justified  by considering  known and  speculated
emission power sources  in SNe and GRBs. The conclusion  is that while cooling
envelope  emission provides  a  natural  explanation for  the  first peak  (as
discussed in  length in \citealt{NakarPiro14}),  all other sources  are either
ruled          out          or          are          highly          unlikely.                        

{\it  Cooling envelope:}  The energy  source of cooling envelope emission  is the  shock that
crosses the  star and  any surrounding  mass, if it exists,  in regions  where the
diffusion time is  longer than the expansion dynamical time.  In these regions
the internal energy deposited by the shock  is trapped by the gas and it cools
adiabatically during  the gas expansion, hence  the term ``cooling  envelope''. As
the outflow  expands its optical  depth drop and  so does the radiation diffusion time to the observer,
while the  expansion time grows. In  regions where the two  time scales become
comparable  the   radiation  escapes   to  the   observer.  As   discussed  in
\cite{NakarPiro14} this source of emission  provides a natural explanation for
the first  optical peak observed on  a time scale of  $\sim$day in case  of a
double peaked SNe. Important supporting evidence for that in the case of SN
2006aj is the  very blue color of the  first peak and the fact  that the colors
do not  vary with  time, which  indicates  that the band that we observe are most likely on the Rayleigh-Jeans part of a blackbody spectrum. This  is expected for  cooling envelope
emission, where the optical depth at the  source is high and the radiation has
enough time  to achieve  thermal equilibrium  before it  escapes, even  if the
outflow  is   fast,  after  the  ejecta   expanded  considerably  \citep{NS10}.

{\it Radioactive decay of $^{56}$Ni:}  This power source of energy, which dominates many SNe light curves, deposits energy at the known decay rate  of $^{56}$Ni,
first to $^{56}$Co and then to $^{56}$Fe. The observed luminosity form
$^{56}$Ni is the energy deposited by radioactive decay in ``exposed'' regions, from where
radiation can escape  over a dynamical time.  Since the total amount of  mass in exposed regions  depends on time and velocity,  the evolution of luminosity generated
by $^{56}$Ni for a given outflow is set by the fraction of $^{56}$Ni in that
region \citep{PiroNakar13}.  This sets, at any given  time, a maximal luminosity 
that  $^{56}$Ni  can  produce, which  is the luminosity of 
an outflow that is composed purly of $^{56}$Ni. The exposed mass at
the first  peak is  $\sim 0.01 (v/0.2c) M_\odot$ implying that  the maximal  contribution of
$^{56}$Ni to the luminosity at this time is $\sim 10^{42} (v/0.2c)$ erg/s.
This  rules out  $^{56}$Ni  as the  source  of  the first  peak  which show  a
luminosity $>3\times 10^{44}$ erg/s.  

{\it Interaction (afterglow):} Another power source seen in some SNe is a continuous  interaction   with  the   circum-stellar  medium. In   GRBs  such
interaction is the source of  the afterglow. The difference between continuous
interaction and cooling  envelope emission is that in the former
the shock  takes place in a region with optical  depth that allows for
radiation to  escape immediately  over a  dynamical time  scale. Thus,  if the
first peak is generated by interaction  then its luminosity is limited by the instantaneous 
strength  of the  interaction. Namely,  the explosion  ejecta must drive  a strong
shock into the circum-stellar medium at least up  to $t \approx 10$ hr, at which point
either the interaction  dies (e.g., due to a sharp  drop in the circum-stellar
density) or  the shock  becomes radiatively inefficient.  The radio  emission at
$t=1.89$  day  is presumably generated  by  such  interaction  and  it  shows  that  the
interaction shock  is propagating at a  velocity close to the  speed of light \citep{Soderberg06}.
The interaction  at this  point is much  too weak to  account for  the optical
emission at this epoch, but assuming that at $t \approx 10$ hr the interaction
have  been much  stronger, could  it  then be  the source  of the  optical/UV?
Considering  all  the  outcomes  of   the  entire  allowed  phase  space  for
interaction  is  beyond the  scope  of  this  paper, however  several  general
considerations show  that it is  highly unlikely that interaction  can produce
the  observed  first optical/UV  peak for two reasons -  it predicts an optical/UV spectrum that is too red and an X-ray flux that is too bright compared to the observations.   

As  the  interaction shock  is  mildly relativistic  its  radius at  the  first  peak is  $r  \sim  10^{15}$ cm.  The
circum-stellar medium  must be  optically thin for Thomson scattering at  this radius,  otherwise the
mildly  relativistic shock  breakout  would  have been  taken  place at  $\sim
10^{15}$ cm, resulting in a much brighter and longer signal in gamma-rays then
observed  (equation \ref{eq:breakout}).  Emission from  optically thin  mildly
relativistic shocks are seen in late stages  of GRBs and in some SNe. In these
cases the shock  is collisionless and it generates strong  magnetic fields and
accelerates electrons  to a power-law distribution.  As a  result the dominant  emission is
synchrotron and  the spectrum  above the self  absorption frequency  (which is
typically in the  radio or mm bands)  is a power-law that is  spread over many
orders of magnitude  in frequency with a specific  flux $F_\nu \propto
\nu^{\alpha}$  with $\alpha  \approx  -1$.  This is  very  different than  the
observed   UV  spectrum   where   $\alpha  \approx   2$, which  requires   the
self-absorption frequency to be $\gtrsim 10^{15}$ Hz. However, even the highest
possible circum-stellar density that is  optically thin for Thomson scattering
at  $\sim  10^{15}$   cm  does  not  bring  the   synchrotron  (or  free-free)
self-absorption  frequency  of  a  mildly  relativistic  shock  into  the  UV.
In addition, the synchrotron power-law spectrum also predicts an
X-ray luminosity that is comparable or larger than the UV luminosity, regardless of the location of the self-absorption frequency. In reality at the time of
the  first peak the  X-ray  luminosity is  fainter  than  the UV  by  two orders  of magnitude. 

{\it Continuous central  engine activity:} The last  power source that is often considered in GRBs and sometimes also in SNe is a
continuous energy supply by a central engine, possibly an accreting black-hole
or  a  long lived magnetar.  The  existence  of  such sources  in  SNe  is  still  rather
hypothetical, while in GRBs there is stronger evidence that the central engine
can be active also  on time scales of hour or  days. Nevertheless, the optical
emission is  highly unlikely to  be powered this way.  The reason is  that the
bulk of the  SN ejecta mass, $\sim  2M_\odot$, lies between the  center of the
explosion and  the observer. The  photons observed  in the first  optical peak
must be generated at larger radius than that of the bulk of the ejecta. If the
energy  from the  central engine  is deposited  first in  the $\sim  2M_\odot$
ejecta it  is converted  to kinetic  and thermal energy  of the  ejecta before
radiated away after the ejecta optical  depth drops, over time scale of weeks.
Thus, similarly  to LGRBs,  the energy  generated by  the central  engine must
``penetrate"  through  the  bulk  of  the  mass  before  being  dissipated  to
optical/UV  photons. Again,  like in  LGRBs, this  may be  done if  the engine
continuously launches  relativistic jets. However, based  on GRB observations,
the expected optical/UV/X-ray emission from relativistic jets suffers from the
same problems  of interaction emission.  It usually show a  power-law spectrum
with $\alpha \ll  2$, which does not fit the  observed optical/UV spectrum and
the faint  X-ray emission. More  importantly, a  relativistic jet must  open a
cavity in the SN ejecta inducing strong spherical asymmetry in the slow moving
material, which is ruled  out by the lack of polarization  and by the spectral
line   profiles    seen   in   the   SN    nebular   phase   \citep{Mazzali07}.

\section{Jet propagation in the core and in the extended material}\label{app:jets} The general physics of a relativistic hydrodynamical
jet  that  propagates in  a  surrounding  medium  is  described at  length  in
\cite{Bromberg11b}. Here we  provide a brief outline of  this system, focusing
on the  time that the engine  must work for  the jet to penetrate  through the
core and through  the extended material. A jet that  propagates in surrounding
media forms a  forward-reverse shock structure at its head.  The high pressure
plasma in  the jet  head spills  sideways continuously  as the  jet propagates
forming  a hot  cocoon  that engulfs  the  jet.  This cocoon  may  or may  not
collimate the  jet, depending on the  jet and the external  medium properties.
Since energy  is leaving the jet  head into the cocoon  continuously, the head
propagation depends on  a continuous supply of energy, which  is injected into
the head  by the  jet via the  reverse shock.  Thus, in order  for the  jet to
propagate a given  distance the engine that  launches the jet must  work for a
duration that is  long enough so a  fresh jet material will  continue to cross
the reverse shock during the entire  propagation. Thus, if the head propagates
up to a  distance $r$ at a velocity  $\beta_h c$ the jet working  time must be:
\begin{equation}
	t_{eng}(r) \approx \frac{r}{c \beta_h}(1-\beta_h)
\end{equation}
where  the  term  $1-\beta_h$  includes  the  relative  velocity  between  the
relativistic jet and the head. This term  is $\approx 1$ for a Newtonian head,
implying that  the engine  working time  is simply  the jet  propagation time.
However, if the  head is relativistic then  by the time that  the engine stops
working the head is at  a distance $\approx c t_{eng}$ from the  center and the last
jet element that was  launched by the engine will catch up  with the head only
after $t_{eng}/(1-\beta_h)$. During  that time the jet will  continue to drive
the head forward. Thus, the engine working time needed for a relativistic head
to    propagate   a    distance   $r$    is   much    shorter   than    $r/c$.

The evolution of the jet is determined by finding the properties of the various components (e.g., head, cocoon, etc.) of that system. 
The propagation velocity of the head is set by the balance of the jet luminosity per unit area into the head and the ram pressure of the ambient medium. It is therefore useful to define a dimensionless parameter which is the  ratio between the energy density of the jet and the rest-mass energy density of the ambient medium \citep{Matzner03}
\begin{equation}
	\tilde{L}=\frac{L_j}{\Sigma_j\rho c^3},
\end{equation}
where $L_j$ is the total jet luminosity, $\Sigma_j$ is the jet cross-section at the head and $\rho$ is the ambient medium density at the head location. The propagation velocity of the jet head is:
\begin{equation}\label{eq:beta_head}
	\beta_h = \frac{1}{1+\tilde{L}^{-1/2}} . 
\end{equation}
Thus, the head is relativistic when $\tilde{L} \gg 1$ and Newtonian for $\tilde{L}\ll 1$. The collimation of the jet depends also on the half opening angle at which the jet is launched, $\theta_0$, where for $\tilde{L} < \theta_0^{-4/3}$ the jet is collimated by the cocoon pressure. The jet collimation affects the value of $\Sigma_j$ and thus also the value of $\tilde{L}$. For a given set of parameters \cite{Bromberg11b} obtain:
\begin{equation} \label{eq:Ltilde}
	\tilde{L}=
    \left\{ \begin{array}{lr}
	\left(\frac{L_j}{\rho t^2 \theta_0^4 c^5}\right)^{2/5} & \tilde{L} < \theta_0^{-4/3}~{\rm(Collimated)}  \\
	&\\
	\frac{L_j}{\rho t^2 \theta_0^2 c^5} & \tilde{L} > \theta_0^{-4/3}~{\rm(Uncollimated)}  \\
	\end{array} \right.
\end{equation}
where $t$ is the time since the jet launching started and $\rho$ is the external density near the jet head location. Equations \ref{eq:beta_head} and \ref{eq:Ltilde} together can be solved to find the jet location at time $t$.

The isotropic equivalent luminosity of a typical GRB jet is $L_{iso} \sim 10^{51}$ erg/s and its opening angle is $\theta_0 \sim 10^o$. The beaming corrected jet luminosity is then $L_j=L_{iso} \theta_0^{2}/2$.
In a massive ($M_{core} > M_\odot$) and compact ($R_{core}\sim 10^{11}$ cm) core the density is high and $\tilde{L} \lesssim 1$, resulting in a Newtonian (or at most a mildly relativistic) collimated jet. Thus, the engine working time must be comparable to the time needed for the jet to cross the core:
\begin{equation}
	t_{eng,core} \sim 7 
	\left(\frac{L_{iso}}{10^{51}{\rm~erg~s^{-1}}}\right)^{-1/3}
	\left(\frac{\theta_0}{10 {\rm~deg}}\right)^{2/3}  
	\left(\frac{R_{core}}{10^{11}{\rm~cm}}\right)^{2/3} 
	\left(\frac{M_{core}}{10M_\odot}\right)^{1/3} {\rm~s}.
\end{equation}
Here we used the approximation $\beta_h \approx \tilde{L}^{1/2}$ which is appropriate for Newtonian heads.

The density of the extended material is much lower than in the core. As a result, for a typical GRB jet $\tilde{L}>\theta_0^{-4/3}$, resulting in an uncollimated relativistic jet. The minimal engine working time for a successful jet penetration is shorter than the extended material light crossing time, but it is still much longer than the time it takes the jet to cross the core: 
\begin{equation}
	t_{eng,ext} \sim 150 \left(\frac{L_{iso}}{10^{51} {\rm ~erg~s^{-1}}}\right)^{-1/2} \left(\frac{R_{ext}}{3\times10^{13} {\rm cm}}\right)^{1/2} \left(\frac{M_{ext}}{10^{-2} M_\odot}\right)^{1/2} {\rm~s}.
\end{equation}
where we used the approximation for a relativistic head $\gamma_h \approx \tilde{L}^{1/4}/\sqrt{2}$, where $\gamma_h$ is the head Lorentz factor.

\end{document}